\documentclass{emulateapj}

\usepackage{rotate}
\usepackage{longtable}
\usepackage{color}

\shorttitle{\emph{Chandra} Legacy Survey of the UDS Field}
\shortauthors{Kocevski et al.}

\begin{document}

\title{\emph{X-UDS:} The \emph{Chandra} Legacy Survey of the UKIDSS Ultra Deep Survey Field}

\author{Dale D.~Kocevski\altaffilmark{1}, 
Guenther Hasinger\altaffilmark{2}, Murray Brightman\altaffilmark{3}, Kirpal Nandra\altaffilmark{4}, Antonis Georgakakis\altaffilmark{4}, Nico Cappelluti\altaffilmark{5}, Francesca Civano\altaffilmark{6}, Yuxuan Li\altaffilmark{1}, Yanxia Li\altaffilmark{2}, James Aird\altaffilmark{7}, David M.~Alexander\altaffilmark{7}, Omar Almaini\altaffilmark{8}, Marcella Brusa\altaffilmark{9,10}, Johannes Buchner\altaffilmark{4,11}, Andrea Comastri\altaffilmark{10}, Christopher J.~Conselice\altaffilmark{8}, Mark A.~Dickinson\altaffilmark{12}, Alexis Finoguenov\altaffilmark{4,13}, Roberto Gilli\altaffilmark{10}, Anton M.~Koekemoer\altaffilmark{14}, Takamitsu Miyaji\altaffilmark{15}, James R.~Mullaney\altaffilmark{16}, Casey Papovich\altaffilmark{17,18}, David Rosario\altaffilmark{7}, Mara Salvato\altaffilmark{4}, John D.~Silverman\altaffilmark{19}, Rachel S.~Somerville\altaffilmark{20,21}, Yoshihiro Ueda\altaffilmark{22}}

\affil{$^1$ Department of Physics and Astronomy, Colby College, Waterville, ME 04961, USA; dale.kocevski@colby.edu\\
       $^2$ Institute for Astronomy, University of Hawaii, Honolulu, HI, USA\\
  $^3$ Department of Astronomy, California Institute of Technology, Pasadena, CA, USA\\
  $^4$ Max-Planck-Institut f\"ur extraterrestrische Physik, Garching, Germany\\
  $^5$ Department of Astronomy, Yale University, New Haven, CT, USA\\
  $^6$ Harvard-Smithsonian Center for Astrophysics, Cambridge, MA, USA\\
  $^7$ Department of Physics, Durham University, Durham, UK\\
  $^8$ School of Physics \& Astronomy, University of Nottingham, Nottingham, UK\\
  $^9$ Department of Physics \& Astronomy, University of Bologna, Bologna, Italy\\
  $^{10}$ INAF Osservatorio Astronomico di Bologna, Bologna, Italy\\
  $^{11}$ Millenium Institute of Astrophysics, Santiago, Chile\\
  $^{12}$ National Optical Astronomy Observatories,Tucson, AZ, USA\\
  $^{13}$ Department of Physics, University of Helsinki, Helsinki, Finland\\
  $^{14}$ Space Telescope Science Institute, Baltimore, MD, USA\\
  $^{15}$ Instituto de Astronom\'ia sede Ensenada, Universidad Nacional Aut\'onoma de M\'exico, Mexico\\
  $^{16}$ Department of Physics \& Astronomy, University of Sheffield, Sheffield, UK\\
  $^{17}$ George P.~\& Cynthia Woods Mitchell Institute for Fundamental Physics \& Astronomy, College Station, TX\\ 
  $^{18}$ Department of Physics \& Astronomy, Texas A\&M University, College Station, TX\\
  $^{19}$ Kavli Institute for the Physics \& Mathematics of the Universe, The University of Tokyo, Kashiwa, Japan\\ 
  $^{20}$ Department of Physics and Astronomy, Rutgers, The State University of New Jersey, Piscataway, NJ, USA\\
  $^{21}$ Center for Computational Astrophysics, Flatiron Institute, New York, NY, USA\\
  $^{22}$ Department of Astronomy, Kyoto University, Kyoto, Japan\\
}

\begin{abstract}

  We present the X-UDS survey, a set of wide and deep \emph{Chandra} observations of the Subaru-XMM Deep/UKIDSS Ultra Deep Survey (SXDS/UDS) field.  The survey consists of 25 observations that cover a total area of 0.33 deg$^{2}$.  The observations are combined to provide a nominal depth of $\sim$600 ksec in the central 100 arcmin$^{2}$ region of the field that has been imaged with \emph{Hubble}/WFC3 by the CANDELS survey and $\sim$200 ksec in the remainder of the field.  In this paper, we outline the survey's scientific goals, describe our observing strategy, and detail our data reduction and point source detection algorithms.  Our analysis has resulted in a total of 868 band-merged point sources detected with a false-positive Poisson probability of $<1\times10^{-4}$.  In addition, we present the results of an X-ray spectral analysis and provide best-fitting neutral hydrogen column densities, $N_{\rm H}$, as well as a sample of 51 Compton-thick active galactic nucleus candidates.  Using this sample, we find the intrinsic Compton-thick fraction to be 30-35\% over a wide range in redshift ($z=0.1-3$), suggesting the obscured fraction does not evolve very strongly with epoch.  However, if we assume that the Compton-thick fraction is dependent on luminosity, as is seen for Compton-thin sources, then our results are consistent with a rise in the obscured fraction out to $z\sim3$.  Finally, an examination of the host morphologies of our Compton-thick candidates shows a high fraction of morphological disturbances, in agreement with our previous results.  All data products described in this paper are made available via a public website.

\end{abstract}


\section{Introduction}

The deepest \emph{ROSAT}, \emph{Chandra} and \emph{XMM-Newton} surveys have resolved the majority of the cosmic X-ray background (CXB) into faint Active Galactic Nuclei (AGN) at $z<5$ and have revolutionized our view of the accretion history of the universe (see, e.g., Brandt \& Hasinger 2005).  However, several open issues remain in our understanding of supermassive black hole (SMBH) growth and its relationship to the evolution of galaxies.  Among these is the uncertain nature of the first accreting black holes (BHs) at ``Cosmic Dawn'' ($z>6$).  The tension between the need for the efficient and rapid accretion required by the existence of SMBH already at $z>7$ and the strict upper limit on their integrated emission from the CXB (Salvaterra et al.~2012) indicates that BHs are rare in high-redshift galaxies or that accretion is heavily obscured.  Another missing piece of the growing BH puzzle is the prevalence of heavily obscured, ``Compton-thick'' AGN (CTAGN) at “Cosmic Noon” ($z\sim2$), when SMBH growth is at its peak.  It is during this obscured phase that SMBHs are predicted to accrete the bulk of their mass and produce most of their feedback into their host galaxies (Hopkins et al.~2006).  Analysis of the CXB requires this population (Worsley et al.~2005; Gilli et al.~2007), but the fraction of AGN that are heavily obscured and the demographics of their host galaxies are still uncertain.


In this paper, we present a wide and deep \emph{Chandra} survey of the Subaru-XMM Deep/UKIDSS Ultra Deep Survey (SXDS/UDS) field, that is designed to help shed light on SMBH growth in two key epochs: Cosmic Dawn at $z>6$ and Cosmic Noon at $z\sim2$.   This survey, hereafter referred to as the X-UDS survey, has two main scientific goals:  (1) We plan to use the survey's deep observations to identify a sizable number of CTAGN via their X-ray spectral signatures up to $z\sim2-3$ and determine their obscuration and host properties.  (2) We also aim to extract information on the nature of the first luminous accreting BHs in the universe by cross-correlating large-scale fluctuations in the CXB and the cosmic infrared background (CIB).  This will provide a unique insight into populations of the early BH seeds and galaxies that are inaccessible to current direct studies and yield information of fundamental importance to cosmology. 


The X-UDS survey targets a field that is rich in multiwavelength coverage, including some of the deepest \emph{Hubble}, \emph{Spitzer}, and \emph{Herschel} observations ever taken.  In addition, this field was previously observed by \emph{XMM-Newton} (Ueda et al.~2008) and is now the target of an ultradeep Subaru Hyper Suprime-Cam survey (Aihara et al.~2018a; 2018b).  The X-UDS observations described in this paper will help complete \emph{Chandra's} deep survey of the five premier extragalactic survey fields: the \emph{Chandra} Deep Field South and North (CDFS/N; Alexander et al.~2003; Xue et al.~2011, 2016), the Extended Groth Strip (EGS; Laird et al.~2009; Nandra et al.~2015), COSMOS (Elvis et al.~2009; Civano et al.~2012) and now UDS.  These fields are the targets of numerous legacy surveys, including the CANDELS \emph{Hubble} (Grogin et al. 2011; Koekemoer et al. 2011) and \emph{Herschel} (Elbaz et al.~2011; H.~Inami et al.~in prep) projects, the SEDS \emph{Spitzer} Explorer Program (Ashby et al.~2013), and the 3D-HST Legacy Survey (Brammer et al.~2012). The supporting data collected by these programs are essential to fully exploiting \emph{Chandra's} deep observations. Furthermore, these fields will be magnets for future facilities, such as the James Webb Space Telescope and the thirty-meter class telescopes that will come online over the next decade.


This is the first paper in a series that will present the basic results from the X-UDS survey.  In this paper, we set forth the science goals of the survey ($\S2$), describe the survey design and our data reduction procedures ($\S3$), present a catalog of point-like X-ray sources detected in the UDS field ($\S4$), discuss the results of an X-ray spectral analysis aimed at finding obscured AGN ($\S5$), and examine the evolution of the Compton-thick AGN fraction with redshift ($\S6$).   Throughout this paper, we assume a $\Lambda$CDM cosmology with $H_{0}=70$ km s$^{-1}$, $\Omega_{\rm m}=0.27$ and $\Omega_{\rm vac}=0.73$.

\begin{center}
\tabletypesize{\scriptsize}
\begin{deluxetable*}{ccccccc}
\tablewidth{0pt}
\tablecaption{Observation Log}
\tablecolumns{8}
\tablehead{\colhead{Field} & \colhead{ObsID} & \colhead{R.A.}    & \colhead{Decl.}   & \colhead{Start Time} & \colhead{Exposure} & \colhead{Roll Angle} \\  
           \colhead{Name}  & \colhead{}      & \colhead{(J2000)} & \colhead{(J2000)} & (UT)                 & (ks)               & (deg) }
\startdata

XUDS-1    &  17287  &  02 16 49.97  &  -05 15 59.38  & 2015-09-23 11:25:22  &  47.46  &   83.2  \\
XUDS-2    &  17288  &  02 17 11.84  &  -05 15 54.28  & 2015-10-02 09:06:59  &  48.59  &   74.6  \\
XUDS-3    &  17289  &  02 17 34.06  &  -05 15 59.49  & 2015-09-25 14:55:15  &  45.97  &   80.2  \\ 
XUDS-4    &  17290  &  02 17 59.06  &  -05 15 49.09  & 2015-09-30 21:27:08  &  47.46  &   76.0  \\ 
XUDS-5    &  17291  &  02 18 16.41  &  -05 15 49.03  & 2015-10-04 11:32:14  &  49.43  &   70.2  \\
XUDS-6    &  17292  &  02 16 49.98  &  -05 15 35.73  & 2015-09-24 01:07:34  &  49.14  &   81.8  \\
XUDS-7    &  17293  &  02 17 11.50  &  -05 15 35.80  & 2015-09-28 13:15:49  &  49.43  &   77.4  \\  
XUDS-8    &  17294  &  02 17 34.41  &  -05 15 41.01  & 2015-10-03 13:28:14  &  49.43  &   72.2  \\  
XUDS-9    &  17295  &  02 17 59.05  &  -05 15 35.80  & 2015-09-27 03:54:20  &  46.03  &   78.9  \\
XUDS-10   &  17296  &  02 18 16.75  &  -05 15 35.75  & 2015-09-07 03:16:33  &  49.33  &   90.2  \\    
XUDS-11   &  17297  &  02 16 49.64  &  -05 10 14.28  & 2015-09-08 03:37:11  &  49.34  &   93.4  \\ 
XUDS-12   &  17298  &  02 17 11.50  &  -05 10 14.36  & 2015-09-09 22:58:07  &  49.34  &   92.5  \\ 
XUDS-13   &  17299  &  02 17 34.41  &  -05 10 19.57  & 2015-09-10 16:47:06  &  49.31  &   92.0  \\ 
XUDS-14   &  17300  &  02 17 59.40  &  -05 10 19.54  & 2015-09-13 14:28:13  &  49.40  &   95.2  \\ 
XUDS-15   &  17301  &  02 18 16.40  &  -05 10 14.30  & 2015-09-15 01:48:40  &  49.90  &   90.2  \\ 
XUDS-16   &  17302  &  02 16 49.65  &  -05 05 13.58  & 2015-09-15 16:18:43  &  49.62  &   85.2  \\ 
XUDS-17   &  17303  &  02 17 11.51  &  -05 05 13.65  & 2015-09-18 13:40:58  &  51.19  &   82.2  \\ 
XUDS-18   &  17304  &  02 17 33.71  &  -05 05 13.68  & 2015-07-05 16:10:06  &  44.69  &  105.2  \\ 
XUDS-19   &  17305  &  02 17 58.70  &  -05 05 18.84  & 2015-07-06 20:26:22  &  48.48  &  103.2  \\ 
XUDS-20   &  17306  &  02 18 16.40  &  -05 05 13.60  & 2015-07-08 07:14:00  &  50.79  &  103.7  \\
XUDS-21   &  17307  &  02 16 49.65  &  -05 05 31.40  & 2015-07-09 06:25:36  &  50.81  &  104.2  \\
XUDS-22   &  17308  &  02 17 11.51  &  -05 05 31.47  & 2015-07-10 23:29:00  &  44.79  &  106.2  \\
XUDS-23   &  17309  &  02 17 33.72  &  -05 05 31.50  & 2015-09-19 04:25:56  &  51.08  &   79.2  \\
XUDS-24   &  17310  &  02 17 58.35  &  -05 05 36.66  & 2015-08-27 18:49:47  &  47.41  &   98.8  \\
XUDS-25   &  17311  &  02 18 16.39  &  -05 05 41.79  & 2015-09-05 23:34:24  &  48.78  &   94.7  \\

\vspace*{-0.075in}
\enddata
\end{deluxetable*}
\end{center}

\vspace{-0.3in}

\section{Science Goals}

The X-UDS survey was selected in Cycle 16 as a \emph{Chandra} X-ray Visionary Project (XVP) with the goal of investigating several high-impact science questions that could not be easily addressed within the standard time-allocation process. As such, the survey is designed to facilitate a wide range of compelling scientific investigations. In this section, we outline the two main science drivers behind the survey, which divide naturally into two epochs: Cosmic Dawn at $z>6$ and Cosmic Noon at $z\sim2$. 


\subsection{Obscured SMBH Growth at Cosmic Noon}

The first scientific goal of X-UDS is to identify a sizable number of heavily obscured AGN and determine their obscuration and host galaxy properties in the intrinsic luminosity range $L_{\rm X}\sim10^{44}$ to $10^{45.5}$ erg s$^{-1}$ up to $z\sim2$, covering the peak activity of SMBH growth in the Universe.  Population synthesis analyses of the CXB require a substantial fraction of heavily obscured, CTAGN; however, their cosmic fraction and the demographics of their host galaxies are still uncertain (e.g., Comastri et al.~1995; Worsley et al.~2005; Gilli et al.~2007; Akylas et al.~2012; Ueda et al.~2014).  These AGN represent a key phase in the life cycle of galaxies as the majority of SMBH growth is expected to be enshrouded in obscuring gas and dust (Hopkins et al.~2006).
CTAGN are challenging to detect given the strong absorption of their X-ray signals (by factors of 10-100) at energies $<10$ keV.  However, even the most obscured objects with $N_{\rm H}>10^{24}$ cm$^{-2}$ can be identified by sensitive X-ray spectroscopy due to nuclear emission that is Compton-scattered into our line of sight.  This ``reflected'' emission has a characteristic spectral shape consisting of a flat continuum and a high equivalent width Fe K$\alpha$ fluorescence line (Matt et al.~1996).  Identifying obscured AGN becomes easier at higher redshifts as key spectral features associated with CTAGN, such as the Fe K$\alpha$ line at 6.4 keV and the Compton reflection bump which peaks at 30 keV, redshift into \emph{Chandra}'s 2-8 keV band at $z\sim2$.  Recently, Brightman et al.~(2014) identified 100 CTAGN in the \emph{Chandra} Deep Field South, EGS, and COSMOS fields using spectral models from Brightman \& Nandra (2011), which correctly account for emission from Compton scattering, the geometry of the absorbing material, and include a self-consistent treatment for Fe K$\alpha$ emission. 


To facilitate the identification of CTAGN out to $z\sim2$, the exposure pattern of the X-UDS observations are designed in a way to achieve both a deep coverage ($\sim600$ ksec) over the central CANDELS region and wide coverage to facilitate the CXB/CIB fluctuation work.  The UDS field has deep \emph{Hubble}, \emph{Spitzer}, and \emph{Herschel} observations, allowing us to determine the morphologies, masses, and star formation rates (SFR) of AGN hosts as a function of obscuration to $z\sim3$.  The X-UDS observations, in conjunction with the extensive multiwavelength data already available in the field, will allow us to address (1) the evolution of the CTAGN fraction with redshift and (2) the mechanisms that trigger obscured SMBH growth over cosmic time.

Regarding the CTAGN fraction, recent progress in determining the evolution of the AGN hard X-ray luminosity function (Vito et al.~2014; Aird et al.~2015) and population synthesis models of the X-ray background spectrum (Ueda et al.~2014), as well as new spectral analysis models (Brightman \& Nandra 2011), has allowed us to start exploring the redshift-luminosity evolution of CTAGN fraction.  However, the statistical and systematic uncertainties of the current samples are still substantial.  Several studies have reported an increase in the fraction of heavily obscured AGN with redshift (e.g., Hasinger 2008), but the extent of this increase is still debated (Treister et al.~2009; Brightman \& Ueda 2012; Buchner et al.~2015). Populating the CTAGN demographics with more data and a better handle on systematic selection effects will be key to improving our understanding of how the CTAGN population evolves with redshift.  We estimate that combining the intrinsically bright CTAGN detected in EGS and UDS with fainter sources from the ultradeep CDFS/N data will allow for a robust determination of the CTAGN X-ray luminosity function (XLF) out to $z=3$ in at least three redshift bins.


Regarding the triggering mechanisms of obscured SMBH growth, galaxy mergers have long been proposed as a means to fuel AGN activity (e.g., Hopkins \& Hernquist 2006), yet studies of X-ray selected AGN out to $z\sim2$ have failed to find the predicted growth of dark matter host halos (Allevato et al.~2011) or the morphological signatures indicative of recent merger activity (Schawinski et al.~2011; Kocevski et al.~2012; Rosario et al.~2015).  However, past studies have not been sensitive to CTAGN due to their extremely attenuated X-ray emission.  Hydrodynamical merger simulations predict that an obscured AGN phase should coincide with the most morphologically disturbed phase of a galaxy interaction (Cattaneo et al.~2005; Hopkins et al.~2008). It is therefore likely that many studies have systematically missed the AGN-merger connection by not sampling the obscured AGN population well. We aim to use the existing CANDELS imaging in the UDS to search for signs of disturbed host morphologies and recent merger activity among the CTAGN population.  Based on existing samples of CTAGN in CDFS, EGS, and COSMOS, Kocevski et al.~(2015) recently reported a $3.8\sigma$ excess of disturbed morphologies among the heavily obscured AGN at $z\sim1$ compared to unobscured AGN with similar X-ray luminosities.  This is part of an increasing body of work that now connects obscured AGN with galaxy interactions (e.g., Koss et al.~2011, Glikman et al.~2015, Donley et al.~2017, Goulding et al.~2017, Ricci et al.~2017).  The X-UDS observations will increase the sample of CTAGN that have been imaged with \emph{Hubble}/WFC3 and allow for this work to be extended to $z\sim2$ for the first time, where mergers are predicted to play an even greater role in triggering obscured AGN.


\subsection{Fingerprints of SMBH at Cosmic Dawn}

The CIB is the collective radiation emitted throughout cosmic history as observed at infrared wavelengths.  The intensity, spectrum, and spatial fluctuations of the CIB all provide valuable information about sources too faint to be directly detected (e.g., Kashlinsky et al.~2005, 2007, 2012; Cooray et al.~2012; Helgason et al.~2012).  A salient feature of the CIB fluctuations is that their spatial power spectrum rises a factor of 10 above the expected contribution from all known sources at angular scales larger than $20^{\prime\prime}$ (Kashlinsky et al.~2012).  It has been proposed that these fluctuations in the CIB arise from objects within the first 0.5 Gyr of the universe ($z>6$).  This is evidenced by the fact that there are no large-scale correlations between the source-subtracted \emph{Spitzer}/IRAC imaging and faint \emph{Hubble}/ACS sources in the GOODS regions observed to $m_{\rm AB}\sim28$ (Kashlinsky et al.~2007). This implies that unless the CIB anisotropies arise in more local but extremely faint and so far unobserved galaxies (i.e., $m_{\rm AB}>28$), the sources responsible for the near-infrared emission must be redshifted beyond $z=6$, putting the Lyman break redward of the longest ACS wavelength ($\sim9000$\AA).  The CIB fluctuations are also isotropic, described by the $\Lambda$CDM model density field at high $z$; this clustering is significantly different from that of known galaxy populations at recent epochs.  Similar results have been obtained with \emph{AKARI} by Matsumoto et al.~(2011), who showed that the spectral energy distribution (SED) of the fluctuations has a roughly $\lambda^{-3}$ dependence out to 2.4 $\mu$m, which is consistent with the superposition of Rayleigh-Jeans spectra from high-$z$ sources.

\begin{figure}[t]
\epsscale{1.1}
\plotone{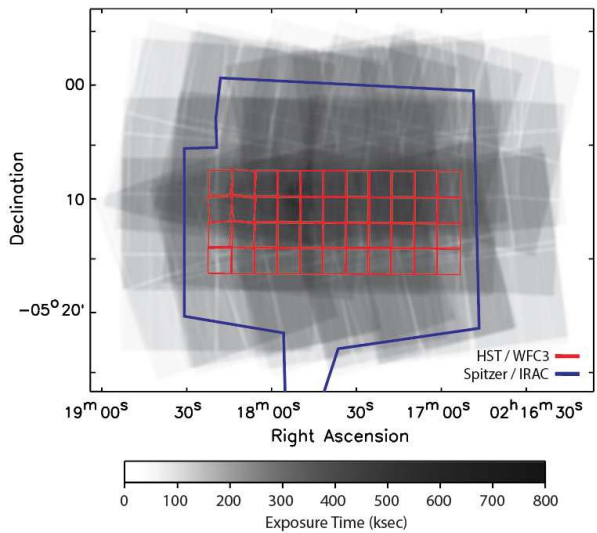}
\caption{Combined exposure map for the X-UDS observations.  The mosaic is a combination of 25 observations that have an average exposure of 50 ksec each.  The outer blue box denotes the region with deep \emph{Spitzer} IRAC observations taken by the SEDS survey.  The inner red outline shows the region imaged with \emph{Hubble}/WFC3 by the CANDELS survey.  \label{fig-expmap} }  
\vspace*{0.1in}
\end{figure}

Recently, Cappelluti et al.~(2013) measured a significant cross-correlation signal between the source-subtracted CIB and CXB fluctuations in the Extended Groth Strip. The \emph{Chandra} 0.5-2 keV unresolved CXB fluctuations on scales $20^{\prime\prime} - 800^{\prime\prime}$ are highly coherent with the CIB, with an overall significance of $\sim3.8\sigma$ and $\sim5.6\sigma$, for the 3.6 and 4.5$\mu$m IRAC bands, respectively, suggesting significant active BH populations among the CIB sources. The measured cross-power indicates that objects associated with powerful X-ray emitters produce 15-25\% of the CIB power. The coherence of the CXB-CIB cross-correlation signal is well above that from known sources (Helgason et al.~2012), and it is therefore likely that a substantial growing BH population contributes a large fraction of the CIB in the early universe (e.g., Mirabel et al.~2011).

If the sources responsible for the CXB and CIB fluctuations are at high redshift (i.e., $z>6$) and distributed according to a $\Lambda$CDM density field, their angular CXB/CIB fluctuation spectrum should dominate in the region around $1000^{\prime\prime}$ (4 Mpc), depending on the epoch of the sources.  Thus far, the joint CXB and CIB fluctuations have been studied on scales $<800^{\prime\prime}$ in the EGS field, but, due to its elongated configuration ($8^{\prime}\times 45^{\prime}$), scales exceeding $\sim300^{\prime\prime}$ are poorly sampled.  The UDS field has been observed homogeneously with warm \emph{Spitzer} IRAC observations as part of the SEDS survey that are of depth similar to those in EGS, but in a geometry ($22^{\prime}\times22^{\prime}$) that is much better suited for the determination of the cosmologically interesting signal at large angular scales.

\begin{figure}[t]
\epsscale{1.1}
\plotone{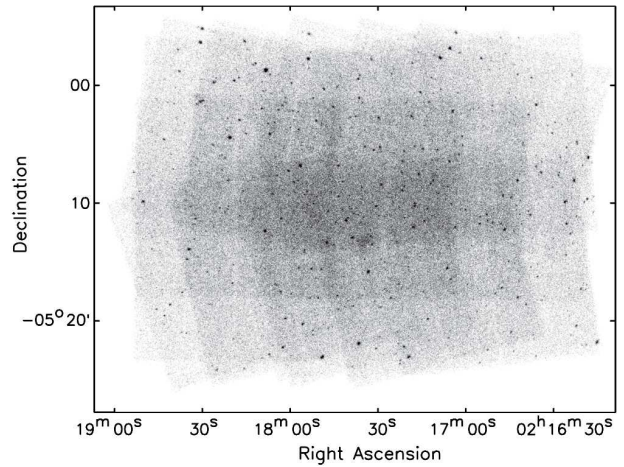}
\caption{\emph{Chandra} full-band (0.5-7 keV) mosaic image of the 25 observations that make up the X-UDS dataset. \label{fig-evtimg} }  
\vspace*{0.1in}
\end{figure}

The X-UDS observations are designed to match the geometry of the existing \emph{Spitzer} observations, allowing for the cross-correlation of the unresolved diffuse CIB and CXB signals.  This will provide a unique insight into populations of the early universe unobtainable by other means, yielding information of fundamental importance to cosmology.  A primary goal of X-UDS is, therefore, to measure the shape of the CXB/CIB fluctuation spectrum, determine whether the large-scale fluctuations are due to high-$z$ sources, and extract information on the first luminous sources in the universe.

%


\begin{figure*}[t]
\epsscale{1.1}
\plotone{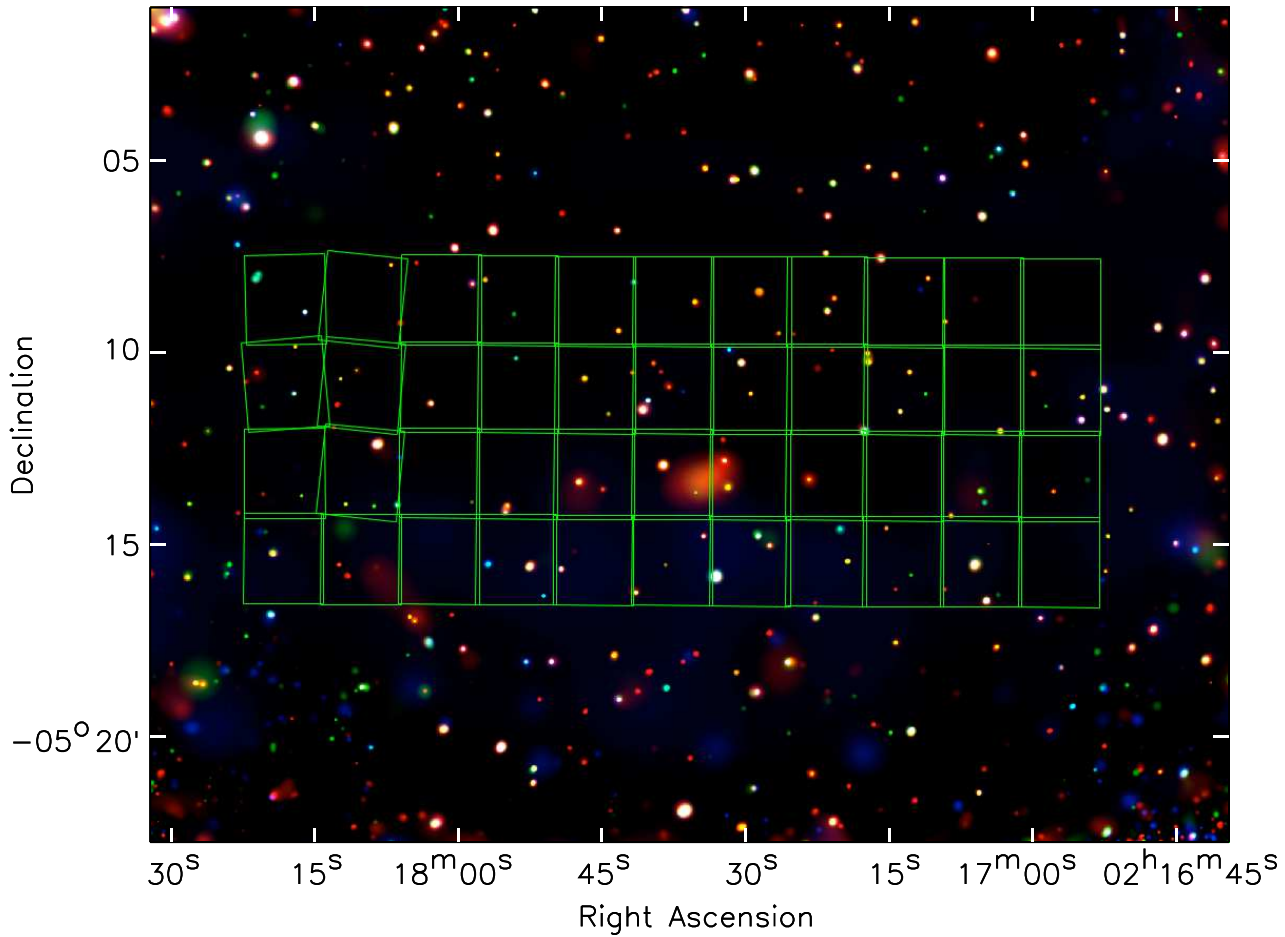}
\caption{Adaptively smoothed ``true-color'' image of the center of the UDS field.  Colors correspond to 0.5-2.0 keV (red), 2-4 keV (green), and 4-8 keV (blue).  The green outline highlights the location of the \emph{Hubble}/WFC3 observations from the CANDELS survey. \label{fig-RGBimg} }  
\vspace*{0.1in}
\end{figure*}

\section{Observations and Data Processing}

\subsection{\emph{Chandra} Observations}

The X-UDS observations were carried out with \emph{Chandra's} Advanced CCD Imaging Spectrometer (ACIS; Garmire et al. 2003) in between the time period 2015 July 6 to 2015 October 4.  The observations consist of 25 pointing positions that cover a total area of roughly $35^{\prime}\times25^{\prime}$ in size.  The exposure times and roll angles of the individual observations range from $70.2^{\circ}$ to $106.2^{\circ}$ and 44.69 to 51.19 ksec, respectively.  A summary of the observational parameters of the X-UDS exposures is listed in Table 1.  Each pointing was imaged with the $16^{\prime}9 \times 16^{\prime}9$ ACIS-I array, with the aim point located on the ACIS-I3 chip. The ACIS-S2 chip was also active during the observations, but due to its large off-axis angle and reduced effective area, we do not make use of it in this analysis.  All observations were carried out in the FAINT telemetry mode with the nominal 3.2 sec CCD frame time.  The individual pointings are mosaicked to provide an average exposure time of 200 ksec over the outskirts of the field and 600 ksec in the central region that overlaps the CANDELS \emph{Hubble}/WFC3 imaging.  An exposure map of the combined X-UDS observations is shown in Figure \ref{fig-expmap}.



\subsection{X-Ray Data Reduction}

The data reduction was performed using the CIAO data analysis software v4.7 (Fruscione et al.~2007), closely following the prescription described in Laird et al.~(2009; L09) and Nandra et al.~(2015).  Briefly, for each individual observation (hereafter ObsID), we corrected the level 1 event files for aspect offsets, applied destreaking algorithms, and identified hot pixels and cosmic-ray afterglows for removal using the {\tt acis\_find\_afterglow} task.  New level 2 event files were then created after correcting for charge transfer inefficiency (CTI) and gain effects using the {\tt acis\_process\_events} task.  To identify periods of anomalously high background, we created light curves for each ObsID in the 0.5-7 keV band with a bin size of 50 s.  This analysis was restricted to ACIS chips 0-3, and \emph{ASCA}-style event grades 0, 2, 4 and 6.  Periods of high background were rejected using the procedure of Nandra et al.~(2005), adopting a threshold of 1.4 times the quiescent background level, determined as the count rate at which the background shows zero excess variance over that expected from statistical fluctuations alone. 



Following this basic reduction, we corrected the astrometry of the individual image frames using a reference catalog.  For this task, we made use of the UKIDSS (Lawrence et al. 2007) DR10 $K$-band catalog\footnote{Available at http://wsa.roe.ac.uk/dr10plus\_release.html} to register the X-UDS images to the near-infrared reference frame.  We first ran the \emph{Chandra} wavelet source detection task {\tt wavdetect} on the 0.5-7 keV image, using a detection threshold of $10^{-6}$, then used the CIAO task {\tt reproject\_aspect} to correct the astrometry compared to the reference image and create new aspect solution files.  The new aspect solutions were then applied to the event files using the task {\tt reproject\_events}.  The parameters used for {\tt reproject\_aspect} were a source match radius of $2^{\prime\prime}$ and a residual limit of $0.50^{\prime\prime}$.  Typically, $\sim40$ sources were matched in each ObsID and used in the reprojection.  The absolute value of the applied offsets at this step was consistently small ($<0\farcs5$).

\begin{figure}[t]
\epsscale{1.05}
\plotone{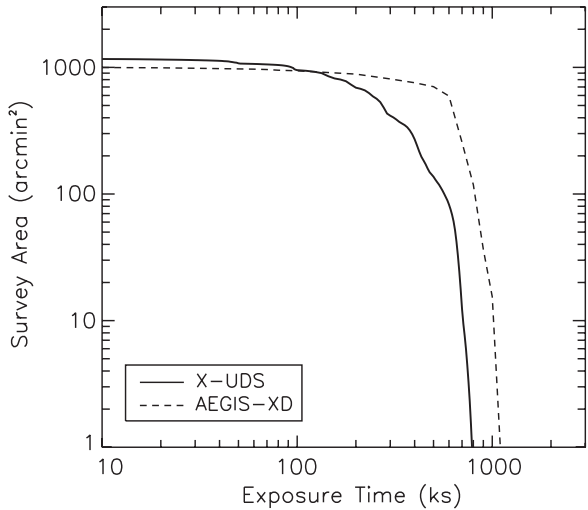}
\caption{Effective exposure time as a function of survey area for the X-UDS observations.  Also shown is the depth of the AEGIS-XD observations for comparison.\label{fig-areaexp} }  
\vspace*{0.1in}
\end{figure}

After this astrometric calibration, we created event files, exposure maps, and point spread function (PSF) maps for each ObsID in the full (FB; 0.5-7 keV), soft (SB; 0.5-2 keV), hard (HB; 2-7 keV) and ultrahard (UB; 4-7 keV) bands.  For the purpose of producing the color mosaic shown in Figure \ref{fig-RGBimg}, we also produced images in the 2-4 keV and 4-8 keV bands.  The exposure maps were created using the task {\tt fluximage} for each of our four primary bands using weights appropriate for a $\Gamma=1.4$ power-law spectrum; these weights are listed in Table 2.  The PSF maps were created using {\tt mkpsfmap} using an enclosed counts fraction of 0.3.  The individual event files were then combined using the {\tt reproject\_obs} task, while the exposure maps were stacked together using the {\tt dmregrid} task.  The individual PSF maps were combined using the task {\tt dmimgfilt} so as to return the minimum PSF value at each location in the combined mosaic.  The final data product of this reduction is a merged event file that covers the entire X-UDS region.  A mosaic image of the merged event file in the full band is shown in Figure \ref{fig-evtimg}.  An adaptively smoothed, ``true-color'' image of the center of the field, created using the {\tt csmooth} algorithm (Ebeling et al.~2006), is shown in Figure \ref{fig-RGBimg}.  The effective exposure time as a function of survey area is shown in Figure \ref{fig-areaexp}.

\subsection{Source Detection and Validation}


Source detection proceeded in the same fashion as that described in Nandra et al.~(2015).  We start by creating a “seed” source catalog using {\tt wavdetect}, which is run with a low false-positive probability threshold ({\tt sigthresh=$10^{-4}$}) in order to capture all potential sources.  Here {\tt wavdetect} was run on the combined event files in our four energy bands with the standard $\sqrt{2}$ set of wavelet scales (i.e., 1, 1.41, 2, 2.82 ... 16).  We also provided {\tt wavdetect} the minimum PSF maps created earlier for each energy band.  The resulting seed catalogs contain 2850, 2021, 2306, and 1625 sources in the full, soft, hard, and ultrahard bands, respectively.
%

We then extract counts from the merged images at the positions of each of our candidate sources.  For the purpose of source validation, counts were measured using a circular aperture with radius equal to the exposure-weighted 50\% encircled energy fraction (EEF) of the \emph{Chandra} PSF.  Here, the PSF size at the location of each candidate source, in each separate ObsID, was taken from a lookup table calculated using the MARX simulation software as described in L09.  An exposure-weighted average PSF size is then computed for each source in the merged image.
%
%
The background near each source was determined using an annulus with an inner radius equal to 1.5 times the 95\% EEF at the source position and an outer radius of 100 pixels larger than this, excluding detected sources.  An average exposure value was also calculated for the source and background areas. The counts in the background area were then scaled to the source region by the ratio of the source and background areas and the ratio of the source and background average exposures, after masking out the 95\% EEF region of other candidate sources.

\begin{center}
\tabletypesize{\scriptsize}
\begin{deluxetable}{ccccc}
\tablecaption{Weights Used for Exposure Map Calculations}
\tablecolumns{5}
\tablehead{\colhead{Energy} & \colhead{Full}          & \colhead{Soft}        & \colhead{Hard}      & \colhead{Ultrahard} \\  
           \colhead{(keV)}  & \colhead{(0.5-7 keV)}   & \colhead{(0.5-2 keV)} & \colhead{(2-7 keV)} & \colhead{(4-7 keV)}  }
\startdata

0.65  &  0.2480  &  0.3867  &    -      &    -     \\
0.95  &  0.1509  &  0.2352  &    -      &    -     \\
1.25  &  0.1042  &  0.1625  &    -      &    -     \\
1.55  &  0.0776  &  0.1209  &    -      &    -     \\
1.85  &  0.0607  &  0.0947  &    -      &    -     \\
2.50  &  0.1359  &    -     &  0.3789   &    -     \\
3.50  &  0.0842  &    -     &  0.2346   &    -     \\
4.50  &  0.0590  &    -     &  0.1645   &  0.4256  \\
5.50  &  0.0445  &    -     &  0.1240   &  0.3208  \\
6.50  &  0.0352  &    -     &  0.0980   &  0.2536  \\

\vspace*{-0.075in}
\enddata
\end{deluxetable}
\end{center} 

\vspace*{-0.4in}
Next the Poisson false probability of observing the total counts measured, given the expected background, was calculated for each source.  A significance threshold of $1\times10^{-4}$ was then applied\footnote{This threshold is unrelated to the {\tt sigthresh} cut used for {\tt wavdetect}.}, and a further detection iteration was performed to mask out only sources more significant than this.  This second iteration ensures that the background is not underestimated due to the masking of random positive variations identified as candidate sources by {\tt wavdetect}.  Any source detected at this $1\times10^{-4}$ probability level in the second iteration in any individual band was included in the final catalog.  As a one-sided {\emph p} value, our threshold of $1\times10^{-4}$ roughly corresponds to a $3.7\sigma$ detection above the expected background.

This process was carried out separately for each of our four energy bands and sources considered significant were band-merged using a matching radius that depends on the exposure-weighted average off-axis angle of the source.  Our adopted cross-band matching radii are $1\farcs{30}$, $2\farcs{44}$, $4\farcs{79}$, and $7\farcs{08}$ for sources with off-axis angles of 0-3, 3-6, 6-9, and $>9$ arcmin, respectively\footnote{Simulations carried out in L09 indicate that positional accuracy is only mildly dependent on source counts, so for the purposes of cross-band matching, we only use off-axis angles to determine our matching radii.}.  These matching radii are roughly 4 times the $1\sigma$ positional uncertainties at these off-axis angles, as determined from MARX simulations (see L09).


Photometry was then performed to estimate source fluxes in several energy ranges, even if the source was not considered a significant detection in that particular band.  For these measurements, we used circular apertures to extract the counts from the combined images, using the exposure-weighted 90\% EEF PSF appropriate for each source.  Fluxes and $1\sigma$ confidence limits were estimated using the Bayesian methodology described in L09, using a spectral slope of $\Gamma = 1.4$ with Galactic $N_{\rm H}$ of $2.54\times10^{20}$ cm$^{−2}$ (Dickey \& Lockman 1990).  Using this method, the best estimate of the source flux is obtained by finding the mode of the posterior distribution function, which is the product of the prior probability distribution and the Poisson likelihood of obtaining the observed total counts for a given source rate and background.  We assumed a prior probability distribution for source fluxes that is based on the observed log $N$-log $S$ relation reported in Georgakakis et al.~(2008).  This approach helps to correct for Eddington bias (Eddington 1940), which causes the flux of faint sources to be generally overestimated using classical approaches.  The resulting fluxes are then extrapolated to the standard energy bands of 0.5-10 keV, 2-10 keV, and 5-10 keV.  All fluxes are on-axis values and corrected for aperture size.

In addition, we also calculated fluxes using the classical method of converting count rates to fluxes.  Effective on-axis source count rates were calculated by dividing the net counts with the average value of the exposure map (in units of count photon$^{-1}$ cm$^{2}$ s) and aperture correcting.  These count rates were converted to fluxes using energy conversion factors of 4.25, 1.71, 8.83, and 1.26 $\times10^{-9}$ energy photon$^{-1}$ in the full, soft, hard, and ultrahard bands, which are appropriate for a $\Gamma = 1.4$ spectrum.   We then extrapolate the fluxes in the full, hard and ultrahard bands out to 10 keV.  Finally, hardness ratios (HRs) were calculated using the Bayesian methodology of Park et al.~(2006).  For this, we employed the BEHR package\footnote{Available at http://http://hea-www.harvard.edu/astrostat/behr/}, which models the detected counts as a Poisson distribution and gives error bars and reliable HRs for sources with both low and high counts.  For comparison, we also calculated HRs using the classical method of ${\rm HR}=(H-S)/(H+S)$, where $H$ and $S$ are the hard and soft band net counts, respectively, corrected to on-axis values.  Sources that only have upper limits on their soft or hard band counts have HR values set to +1 and -1, respectively.

\begin{center}
\tabletypesize{\scriptsize}
\begin{deluxetable}{cccccc}
\tablecaption{Sources Detected in One Band but Not Another}
\tablecolumns{6}
\tablehead{\colhead{Detection} & \colhead{Total number}   & \multicolumn{4}{c}{Nondetection band} \\  
           \colhead{band (keV)}  & \colhead{of sources}   & \colhead{Full} & \colhead{Soft} & \colhead{Hard} & \colhead{Uhrd}}
\startdata

Full (0.5-7)  & 726   &     -    &   203     &   190     &   418     \\
Soft (0.5-2)  & 561   &     62   &    -      &   187     &   317     \\
Hard (2-7)    & 524   &     15   &   154     &    -      &  215       \\
Uhrd (4-7)    & 274   &     12   &    55     &    11     &    -     \\

\vspace*{-0.075in}
\enddata
\end{deluxetable}
\end{center} 

 \vspace*{-0.4in}
\section{Point Source Catalog}

The final band-merged source list in the UDS area consists of 868 sources.  Of these, 726, 561, 524, and 274 were detected at $p<1\times 10^{-4}$ in at least one of the full, soft, hard, and ultrahard bands, respectively.  Sources detected in one band but not another are detailed in Table 3.  The resulting source catalog with full X-ray photometric information is publicly available in FITS table format at http://www.mpe.mpg.de/XraySurveys/, along with supporting data products (i.e., images, event files, and exposure maps).

\subsection{Sensitivity Estimation}

\begin{figure}[t]
\epsscale{1.17}
\plotone{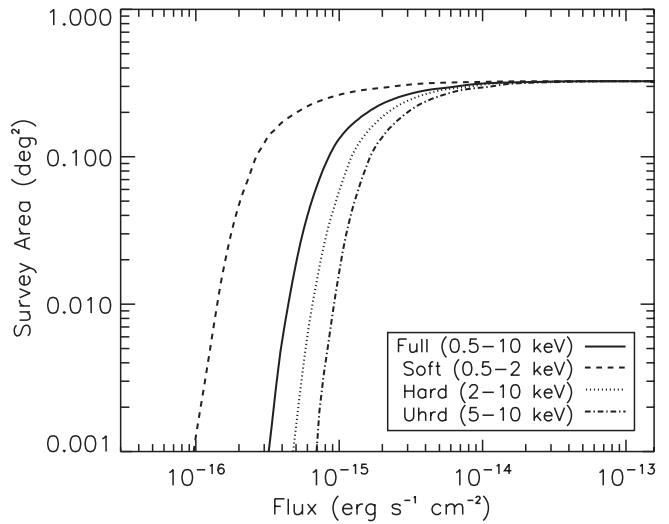}
\caption{Sensitivity curves for the X-UDS survey in the soft (dashed line), hard (dotted line), full (solid line), and ultrahard bands (dot-dashed line), calculated using the methodology of Georgakakis et al.~(2008) \label{fig-areaflux} }  
\vspace*{0.1in}
\end{figure}

To determine the limiting flux of our observations, we have computed sensitivity maps using the method described in Georgakakis et al.~(2008) and implemented more recently in Nandra et al.~(2015).  In short, we begin by computing the average local background per pixel over the entire mosaic in each energy band, after masking out detected sources.  With knowledge of the exposure-weighted PSF size at any location in the image, we then determine the minimum integer number of source counts needed to produce a false-positive Poisson probability that is less than our adopted threshold of $1\times10^{-4}$ given the local background.  This approach accounts for incompleteness and Eddington bias in the sensitivity calculation and is performed in a manner that is also fully consistent with our source detection procedure.  After accounting for the exposure time in the detection cell and the fraction of the total source counts in the cell due to the PSF size, we can determine the minimum flux needed for detection in each band.  Our sensitivity maps are 2D images of this minimum flux over the X-UDS footprint.  The flux limits for the X-UDS survey as a function of area are shown for various energy bands in Figure \ref{fig-areaflux}.  We define the limiting flux of our observations as the flux to which at least 1\% of the survey area is sensitive.  Using this definition, we find the limiting fluxes to be $4.4\times10^{-16}$ (FB; 0.5-10 keV), $1.4\times10^{-16}$ (SB; 0.5-2 keV), $6.5\times10^{-16}$ (HB; 2-10 keV), and $9.2\times10^{-15}$ (UB; 5-10 keV) erg cm$^{-2}$ s$^{-1}$.

\subsection{False Source Estimation}

\begin{center}
\tabletypesize{\scriptsize}
\begin{deluxetable*}{cccccccccccccccc}
\tablecaption{\emph{Chandra} X-UDS Source Catalog: Basic Source Properties}
\tablecolumns{16}
\tablehead{\colhead{ID} & \colhead{CXOUDS} & \colhead{R.A.}      & \colhead{Dec.}  & \colhead{Pos.}  & \colhead{OAA}  & \multicolumn{2}{c}{FB cts}  &
           \multicolumn{2}{c}{SB cts}      & \multicolumn{2}{c}{HB cts}  & \multicolumn{2}{c}{UB cts}  & \colhead{Detection} & \colhead{log $p_{\rm min}$}        \\
           \colhead{}    & \colhead{}    & \colhead{}   & \colhead{}         & \colhead{}          & \colhead{}         & \multicolumn{2}{c}{\line(1,0){30}}   &
           \multicolumn{2}{c}{\line(1,0){30}}  & \multicolumn{2}{c}{\line(1,0){30}}   & \multicolumn{2}{c}{\line(1,0){30}} &  \colhead{}   & \colhead{}    \\
           \colhead{}   & \colhead{}       & \colhead{(J2000)}      & \colhead{(J2000)} & \colhead{Err.}  & \colhead{} & \colhead{$N$}  & \colhead{$B$}  & \colhead{$N$}  &
           \colhead{$B$}  & \colhead{$N$}  & \colhead{$B$}  & \colhead{$N$}  & \colhead{$B$}  & \colhead{Bands}  & \colhead{}  \\
           \colhead{(1)}    & \colhead{(2)}    & \colhead{(3)}   & \colhead{(4)}         & \colhead{(5)}          & \colhead{(6)}         & \colhead{(7)}         &
           \colhead{(8)}        & \colhead{(9)} & \colhead{(10)} & \colhead{(11)} & \colhead{(12)} & \colhead{(13)} & \colhead{(14)} & \colhead{(15)} & \colhead{(16)} }
\startdata
xuds\_001  &  J021749.0-52306  &  34.454190  &  -5.385106  &  0.57  &  8.97  &    862  &  32.5  &  177  &   8.4  &   697  &   25.5  &  294  &   17.4  &  fshu  &  -8.00 \\
xuds\_002  &  J021636.0-52301  &  34.150206  &  -5.383630  &  1.33  &  8.36  &     42  &   6.2  &   22  &   1.6  &    20  &    4.9  &    7  &    3.5  &  fsh   &  -8.00 \\
xuds\_003  &  J021649.0-52237  &  34.204371  &  -5.377063  &  0.96  &  7.50  &     15  &   5.6  &    8  &   1.4  &     7  &    4.3  &    6  &    3.3  &  fs    &  -5.44 \\	
xuds\_004  &  J021736.5-52156  &  34.402224  &  -5.365739  &  0.57  &  8.15  &   1122  &  34.0  &  636  &   9.4  &   484  &   26.5  &  158  &   17.9  &  fshu  &  -8.00 \\
xuds\_005  &  J021637.0-52113  &  34.154250  &  -5.353712  &  1.33  &  8.24  &     56  &  22.5  &   25  &   5.7  &    35  &   18.5  &   17  &   12.3  &  fsh   &  -8.00 \\
xuds\_006  &  J021835.4-52103  &  34.647431  &  -5.350968  &  0.61  &  7.63  &     64  &   9.7  &   41  &   2.5  &    24  &    7.8  &   12  &    5.3  &  fsh   &  -8.00 \\
xuds\_007  &  J021755.7-52017  &  34.481998  &  -5.338286  &  0.44  &  6.58  &    561  &  17.0  &  264  &   4.4  &   297  &   13.2  &  107  &    9.1  &  fshu  &  -8.00 \\
xuds\_008  &  J021712.9-51952  &  34.303808  &  -5.331319  &  0.44  &  6.37  &    277  &  13.9  &  134  &   3.7  &   146  &   10.9  &   43  &    8.0  &  fshu  &  -8.00 \\
xuds\_009  &  J021801.6-51949  &  34.506617  &  -5.330278  &  0.44  &  6.21  &    286  &  14.3  &  165  &   3.6  &   117  &   11.7  &   40  &    7.4  &  fshu  &  -8.00 \\
xuds\_010  &  J021842.8-51934  &  34.678333  &  -5.326295  &  0.61  &  7.97  &     71  &  11.7  &   26  &   3.1  &    43  &    9.1  &   12  &    6.1  &  fsh   &  -8.00 \\
\enddata


\tablecomments{(1) Source identification number, (2) chandra source name, (3) right ascension, (4) declination, (5) positional error in arcseconds, (6) off-axis angle in arcminutes, (7-14) total counts extracted ($N$) and estimated background counts ($B$) in our four analysis bands, (15) bands in which the source was detected, where bands are full (f), soft (s), hard (h), and ultrahard (u), and (16) log of the minimum Poisson false detection probability among our four analysis bands.  Probabilities lower than $10^{-8}$ are listed as -8.0.}

\end{deluxetable*}
\end{center}

\vspace*{-0.325in}

Following the approach of Nandra et al.~(2015), we can constrain the number of spurious detections in our source catalog using our ultrahard band images.  Due to \emph{Chandra's} decreased sensitivity at ultrahard (4-7 keV) energies and the expected spectral shape of the source population, we do not expect sources to be detected exclusively in the ultrahard band without corresponding detections in the full, soft, or hard bands.  Even the most obscured, Compton-thick sources are often detected at soft energies due to the presence of scattered X-ray light (e.g., Brightman \& Nandra 2012).  Therefore, we expect the number of ultrahard-only detections to provide insight on the number of spurious sources in each band.  In our final catalog, there are nine sources detected in the ultrahard band that pass our threshold cut and also lack detections in any other band.  This implies that over our four detection bands, 36 sources may be spurious, or roughly 4.1\% of our final catalog of 868 sources.


\begin{center}
\tabletypesize{\scriptsize}
\begin{deluxetable*}{cccccccccccc}
\tablecaption{\emph{Chandra} X-UDS Source Catalog: Source Fluxes and HRs}
\tablecolumns{12}
\tablehead{\colhead{ID} & \multicolumn{4}{c}{\vspace{-0.05in} Bayesian Flux} & \multicolumn{4}{c}{Classical Flux} & \colhead{Bayes.} & \colhead{Class.} & \colhead{Phot.} \\ 
           \colhead{}    & \multicolumn{4}{c}{\line(1,0){170}}                 & \multicolumn{4}{c}{\line(1,0){170}} & \colhead{}         & \colhead{}          & \colhead{}  \\
           \colhead{}    & \colhead{$F_{0.5-10}$} & \colhead{$F_{0.5-2}$} & \colhead{$F_{2-10}$} & \colhead{$F_{5-10}$} & \colhead{$F_{0.5-10}$} & \colhead{$F_{0.5-2}$} & \colhead{$F_{2-10}$} & \colhead{$F_{5-10}$}  & \colhead{HR} & \colhead{HR} & \colhead{Flag} \\
           \colhead{(1)}    & \colhead{(2)}    & \colhead{(3)}   & \colhead{(4)}         & \colhead{(5)}          & \colhead{(6)}         & \colhead{(7)}         & \colhead{(8)}        & \colhead{(9)} & \colhead{(10)} & \colhead{(11)} & \colhead{(12)}}
\startdata
xuds\_001  &  $109.70^{+4.03}_{-3.89}$  &  $11.89^{+1.03}_{-0.95}$  &  $126.67^{+5.19}_{-5.00}$  &  $81.59^{+5.41}_{-5.10}$   &  $110.04^{+4.03}_{-3.89}$  &  $12.07^{+1.03}_{-0.95}$  &  $127.16^{+5.19}_{-5.00}$  &  $82.38^{+5.41}_{-5.10}$  &  $ 0.51^{+0.03}_{-0.03}$    &  $ 0.51$  & 0 \\
xuds\_002  &  $ 15.72^{+3.10}_{-2.80}$  &  $ 4.86^{+1.24}_{-1.07}$  &  $  8.34^{+2.93}_{-2.52}$  &  $<2.92$		              &  $ 16.54^{+3.48}_{-2.98}$  &  $ 5.28^{+1.50}_{-1.21}$  &  $  9.65^{+3.54}_{-2.83}$  &  $<5.46$		       	    &  $-0.29^{+0.18}_{-0.16}$  &  $-0.30$  & 0 \\
xuds\_003  &  $  2.77^{+1.69}_{-1.54}$  &  $ 1.03^{+0.66}_{-0.52}$  &  $ <1.53$                  &  $<2.26$		              &  $  3.94^{+2.08}_{-1.60}$  &  $ 1.47^{+0.89}_{-0.62}$  &  $<4.03$                   &  $<5.31$		       	    &  $-0.42^{+0.27}_{-0.38}$  &  $-1.00$  & 0 \\
xuds\_004  &  $ 95.53^{+3.04}_{-2.95}$  &  $28.38^{+1.19}_{-1.15}$  &  $ 59.61^{+3.02}_{-2.88}$  &  $28.22^{+2.80}_{-2.58}$   &  $ 95.75^{+3.04}_{-2.95}$  &  $28.50^{+1.19}_{-1.15}$  &  $ 59.96^{+3.02}_{-2.88}$  &  $28.80^{+2.80}_{-2.58}$  &  $-0.24^{+0.03}_{-0.03}$  &  $-0.24$  & 0 \\
xuds\_005  &  $  4.15^{+1.05}_{-0.96}$  &  $ 1.23^{+0.37}_{-0.33}$  &  $  2.50^{+1.24}_{-1.18}$  &  $<1.28$		              &  $  4.49^{+1.14}_{-1.00}$  &  $ 1.38^{+0.43}_{-0.35}$  &  $  3.21^{+1.36}_{-1.14}$  &  $<2.95$		         	&  $-0.19^{+0.21}_{-0.22}$  &  $-0.19$  & 0 \\
xuds\_006  &  $ 12.85^{+2.02}_{-1.86}$  &  $ 4.43^{+0.80}_{-0.72}$  &  $  5.36^{+1.99}_{-1.74}$  &  $<3.57$		              &  $ 13.28^{+2.21}_{-1.95}$  &  $ 4.62^{+0.90}_{-0.76}$  &  $  6.26^{+2.30}_{-1.87}$  &  $<2.76$		       	    &  $-0.42^{+0.13}_{-0.14}$  &  $-0.43$  & 0 \\
xuds\_007  &  $ 43.22^{+1.97}_{-1.89}$  &  $10.06^{+0.68}_{-0.64}$  &  $ 35.32^{+2.29}_{-2.16}$  &  $18.62^{+2.23}_{-2.02}$   &  $ 43.43^{+1.97}_{-1.89}$  &  $10.16^{+0.68}_{-0.64}$  &  $ 35.65^{+2.29}_{-2.16}$  &  $19.16^{+2.23}_{-2.02}$  &  $ 0.01^{+0.04}_{-0.05}$    &  $ 0.01$  & 0 \\
xuds\_008  &  $ 21.83^{+1.47}_{-1.39}$  &  $ 5.24^{+0.51}_{-0.47}$  &  $ 17.50^{+1.73}_{-1.59}$  &  $ 6.79^{+1.36}_{-1.23}$   &  $ 21.96^{+1.47}_{-1.39}$  &  $ 5.30^{+0.51}_{-0.47}$  &  $ 17.86^{+1.73}_{-1.59}$  &  $ 7.18^{+1.56}_{-1.34}$  &  $-0.01^{+0.06}_{-0.06}$  &  $-0.01$  & 0 \\
xuds\_009  &  $ 22.01^{+1.46}_{-1.38}$  &  $ 6.40^{+0.56}_{-0.52}$  &  $ 13.02^{+1.51}_{-1.37}$  &  $ 6.05^{+1.28}_{-1.15}$   &  $ 22.14^{+1.46}_{-1.38}$  &  $ 6.50^{+0.56}_{-0.52}$  &  $ 13.37^{+1.51}_{-1.37}$  &  $ 6.42^{+1.45}_{-1.24}$  &  $-0.25^{+0.06}_{-0.06}$  &  $-0.25$  & 0 \\
xuds\_010  &  $ 14.90^{+2.25}_{-2.08}$  &  $ 2.78^{+0.70}_{-0.61}$  &  $ 12.30^{+2.60}_{-2.23}$  &  $<3.09$		              &  $ 15.37^{+2.45}_{-2.18}$  &  $ 3.00^{+0.81}_{-0.66}$  &  $ 13.38^{+3.00}_{-2.57}$  &  $<4.28$                  &  $ 0.13^{+0.15}_{-0.14}$    &  $ 0.13$  & 0 \\

\enddata

\tablecomments{(1) Source identification number, (2-5) source fluxes in units of $10^{-15}$ erg cm$^{-2}$ s$^{-1}$ in our four analysis bands calculated using the Bayesian methodology of L09, (6-9) source fluxes in units of $10^{-15}$ erg cm$^{-2}$ s$^{-1}$ in our four analysis bands calculated using the classical method of converting count rates to fluxes, (10-11) source hardness ratios calculated using the Bayesian and classical methods, (12) photometric quality flag indicating possible (``1'') and likely (``2'') contamination from a nearby source; all other sources have a flag values of ``0''.  See \S4.3 for details.}


\end{deluxetable*}
\end{center}

\subsection{Catalog Description}

The final X-UDS source catalog of 868 sources is presented in Tables 4 and 5.  Table 4 provides basic source properties such as source position, counts, and detection information.  A more detailed description of each column in Table 4 is reported below.


\begin{enumerate}
\item Column 1: Source identification number
\item Column 2: \emph{Chandra} source name, following the standard IAU convention with prefix ``CXOUDS" for ``\emph{Chandra} X-ray Observatory UDS survey."
\item Columns 3-4: Right ascension and declination in the J2000 coordinate system.
\item Column 5: Positional error in arcseconds.  Based on simulations carried out by L09, the positional error assigned to each source is dependent on its average off-axis angle (OAA) and net counts in the Full band.  These errors range from 0\farcs{14} for bright sources detected at low OAAs (OAA $<4^{\prime}$ and $N>100$ counts) to 1\farcs{33} for faint sources at larger OAAs (OAA $>8^{\prime}$ and $N<50$ counts).
\item Column 6: Exposure-weighted, average off-axis angle in arcminutes.
\item Columns 7-14: Total source counts ($N$) and background counts ($B$) in the four analysis bands.  Counts are given regardless of whether a source was detected in the band.
\item Column 15: List of the bands in which a source is detected with Poisson false detection probability $<1\times10^{-4}$, where the bands are full (F), soft (S), hard (H), and ultrahard (U).
\item Column 16: Log of the minimum Poisson probability among the four analysis bands.  Probabilities lower than $10^{-8}$ are listed as $-8.0$. 
\end{enumerate}

Table 5 presents the flux and HR information for each source in the final catalog.  A more detailed description of each column in Table 5 is reported below.

\begin{figure*}
\epsscale{0.8}
\plotone{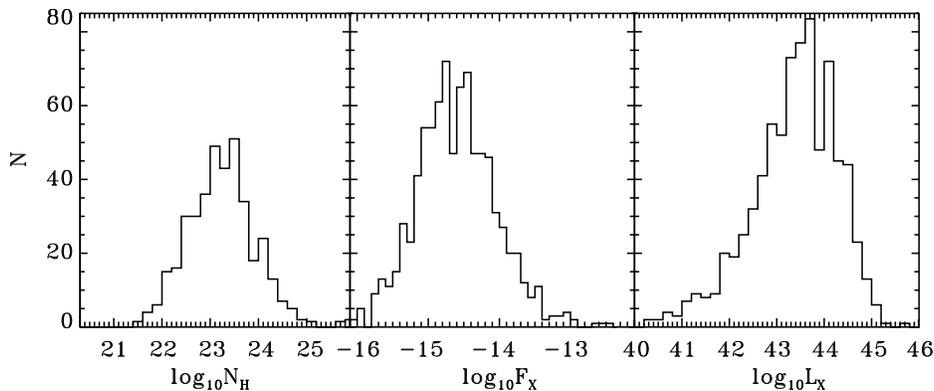}
\caption{Observed distributions of $N_{\rm H}$ (cm$^{-2}$), 0.5$-$8 keV flux (erg cm$^{-2}$ s$^{-1}$) and intrinsic rest-frame 2$-$10 keV luminosity (erg s$^{-1}$) for the X-UDS sources. \label{fig-NHdist}}
\end{figure*}

\begin{enumerate}
\item Column 1: Source identification number
\item Column 2-5: Observed frame source fluxes in our four analysis bands calculated using the Bayesian method described in L09, which corrects for Eddington bias.  Calculations were done using a spectral slope of $\Gamma=1.4$ and Galactic $N_{\rm H}$ of $2.54\times10^{20}$ cm$^{−2}$ (Dickey \& Lockman 1990).  The reported errors are $1\sigma$ values.  Where sources are not detected in a particular band, we report the 68\% upper limit.  All fluxes have units of $10^{-15}$ erg cm$^{-2}$ s$^{-1}$ and have not been corrected for intrinsic source absorption.  
\item Columns 6-9: Observed frame source fluxes in our four analysis bands calculated using the classical method of converting count rates to fluxes.  Calculations were done using a spectral slope of $\Gamma=1.4$ and Galactic $N_{\rm H}$ of $2.54\times10^{20}$ cm$^{−2}$ (Dickey \& Lockman 1990).  Reported errors are $1\sigma$ values.  All fluxes have units of $10^{-15}$ erg cm$^{-2}$ s$^{-1}$ and have not been corrected for intrinsic source absorption. 
\item Column 10-11: Source hardness ratios calculated using the Bayesian and classical methods.  The reported errors on the Bayesian HRs are $1\sigma$ values.  For sources only detected in the full band, and with upper limits in both hard and soft bands, the classical HR cannot be determined and is set to -99.
\item Column 12: Photometry quality flag. A value of ``1" indicates the presence of a nearby source that may be contaminating the photometry. A flag of ``2" indicates that another source was detected within the 90\% EEF and that the photometry is likely heavily contaminated and the source position uncertain. All other sources have a flag of ``0".
\end{enumerate}

\begin{figure*}
\epsscale{1.0}
\plotone{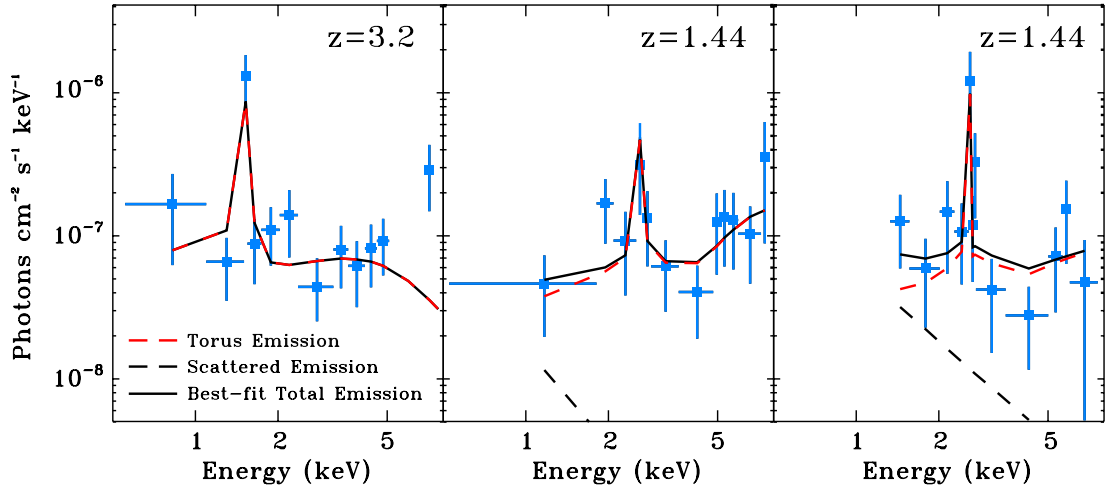}
\caption{X-ray spectra of three Compton-thick AGN detected by our spectra fitting analysis of the X-UDS dataset.  The red dashed line shows the best-fit direct torus emission from the AGN, while the black dashed line shows the Thompson-scattered component.  The solid black line shows the best-fit total emission.  All three sources exhibit strong Fe K$\alpha$ emission characteristic of a Compton-thick AGN. \label{fig-xspec}}
\end{figure*}

\begin{center}
\tabletypesize{\scriptsize}
\begin{deluxetable*}{cccllcc}
\tablewidth{0pt}
\tablecaption{\emph{Chandra} X-UDS Source Catalog: Best-Fit Spectral Parameters}
\tablecolumns{8}
\tablehead{\colhead{ID}   & \colhead{Total Counts} & \colhead{$z$}    & \colhead{log $N_{\rm H}$}   & \colhead{$\Gamma$} & \colhead{log$_{10}$ $F_{\rm 0.5-8}$} & \colhead{log$_{10}$ $L_{\rm 2-10}$}  \\  
           \colhead{(1)}  & \colhead{(2)}          & \colhead{(3)}    & \colhead{(4)}              & \colhead{(5)}      & \colhead{(6)}                  & \colhead{(7)}                }
\startdata

xuds\_001    &    853     &       0.99   &    $22.89^{+0.10}_{-0.12}$  &      $1.22^{+u}_{-0.34}$     &   -12.91    &      44.71  \\
xuds\_002    &     38     &       1.32   &    $20.00$                &      $1.70$         &   -13.85    &      43.93  \\
xuds\_003    &     13     &       2.83   &    $20.00$                &      $1.70^{+u}_{-0.14}$     &    -14.48    &     44.06  \\
xuds\_004    &   1110     &       1.03   &    $20.00$                &      $1.67^{+0.13}_{-0.13}$  &    -13.12    &     44.41  \\
xuds\_005    &     66     &       0.50   &    $22.01^{+u}_{-0.44}$     &      $1.70$        &    -14.36    &     42.57  \\
xuds\_006    &     59     &       0.61   &    $20.00$                &      $1.70$        &    -13.99    &      43.01  \\
xuds\_007    &    576     &       1.30   &    $22.49^{+0.16}_{-0.14}$  &      $1.70$        &    -13.43    &      44.46  \\
xuds\_008    &    281     &       0.81   &    $21.92^{+0.46}_{-0.26}$  &      $2.65^{+u}_{-0.52}$     &    -13.76    &     43.62  \\
xuds\_009    &    303     &       1.95   &    $20.00$                &      $1.70$        &    -13.76    &     44.42  \\
xuds\_010    &     66     &       1.98   &    $22.75^{+u}_{-0.40}$     &      $1.70$        &    -13.91    &     44.40  \\
\vspace*{-0.075in}
\enddata
\tablecomments{(1) Source identification number. (2) Total counts extracted. (3) Redshift for optical counterpart. (4) Best-fit neutral Hydrogen column density in units of log$_{10}(N_{\rm H}/{\rm cm}^{-2})$, with associated 90\% limits.  Errors of ``u'' and ``l'' denote that the uncertainty calculation hit the hard lower and upper limits of the model, which are $10^{20}$ cm$^{-2}$ and $10^{26}$ cm$^{-2}$, respectively.  (5) Best-fit power-law index and associated 90\% limits.  Errors of ``u'' and ``l'' denote that the uncertainty calculation hit the hard lower and upper limits of the model, which are 1.0 and 3.0, respectively. (6) Observed 0.5-8.0 keV flux in units of log$_{10}(F_{\rm X}/{\rm erg~cm}^{-2}~{\rm s}^{-1})$.  (7) Absorption-corrected, 2-10 keV rest-frame luminosity in units of log$_{10}(L_{\rm X}/{\rm erg~s}^{-1})$.}


\end{deluxetable*}
\end{center} 


\section{X-Ray Spectral Fitting}



One of the goals of the X-UDS survey is to better contrain the prevalence of heavily obscured, Compton-thick AGN (CTAGN) versus redshift.  To this end, we have extracted and fitted the {\it Chandra} spectra of all 868 sources in our catalog using the method described in Brightman et al.~(2014; hereafter B14) in order to estimate the nuclear obscuration present in each source.  Individual source spectra were extracted from the processed data using the {\tt acis\_extract} (AE) software version 2014-08-29 (Broos et al.~2010)\footnote{The {\tt acis\_extract} software package is available at http://www.astro.psu.edu/xray/acis/acis\_analysis.html} utilizing CIAO v4.7, MARX v5.1.0 and CALDB v4.6.5.  Events were extracted from regions where 90\% of the PSF has been enclosed at 1.5 keV.  Background spectra were extracted from an events list which has been masked of all detected point sources in that particular field, using regions which contain at least 100 counts.  The data from each ObsID were then merged to create a single source spectrum, background spectrum, RMF and ARF for each source.


The source spectra were grouped with the {\tt Heasoft} v6.16 tool {\tt grppha} with a minimum of one count per bin.  We carried out the spectral fitting using {\tt XSPEC} v12.8.2 on background-subtracted spectra with a modifed version of the Cash fit statistic ($C$-stat, Cash 1979; Arnaud 1996) in the energy range 0.5$-$8 keV. The redshift of each X-ray source was obtained from the best near-infrared counterpart in the CANDELS $H$-band catalog (Galametz et al.~2013) or UKIDSS DR10 $K$-band catalog and are a combination of spectroscopic and photometric redshifts.  Counterpart matching was done using the likelihood ratio technique of Sutherland \& Saunders (1992), following the procedure described by Civano et al.~(2012).  Further details of this matching will be provided in a forthcoming paper (G.~Hasinger et al., in prep.).  For sources without a redshift available, we set $z=0$, likewise for stars.  


As our baseline model, we fitted a simple power law to each spectrum, where the power law index, $\Gamma$, is fixed to 1.7 with only the normalization as a free parameter. We fix $\Gamma$ at 1.7 because this is where the distribution of $\Gamma$ was found to peak in B14. We then tested for the presence of absorption along the line of sight using the {\tt sphere} model of Brightman \& Nandra (2011, BN11), which predicts the power-law spectrum having undergone reprocessing by photoelectric absorption, Compton scattering, and Fe K fluorescence by a spherical distribution of matter surrounding the central X-ray source. In this case, the free parameters are the normalization and $N_{\rm H}$. In order for this model to be chosen as the best-fit model over the simple power law, it must yield an improvement in the fit statistic of $\Delta C$-stat$\geq2.7$, which corresponds to a better fit at $\sim90$\% confidence (see B14 for details). 


Next we tested for the presence of reflection using the {\tt torus} model from BN11, which is identical to the {\tt sphere} model but with a biconical void to simulate a torus-like structure where X-ray photons can scatter within the cone into the line of sight. We tested two scenarios here, where the torus has a half opening angle of 30 degrees and 60 degrees. We also added a secondary power-law component, not subjected to absorption, that represents intrinsic emission Thompson-scattered into the line of sight. We did not allow the ``scattered fraction'' ($f_{\rm scatt}$) to exceed 10\%\footnote{An upper limit of 10\% on the scattered fraction is applied since it would be challenging to identify Compton-thick sources with such strong scattered emission. Furthermore, in the local Universe, the scattered fraction is typically only a few percent (e.g., Noguchi et al.~2010).}.  In this case, the free parameters are the normalization, $N_{\rm H}$, and $f_{\rm scatt}$.  In order for one of these models to be chosen as the best-fit model, it must yield a further improvement in the fit statistic of $\Delta C$-stat$\geq2.7$ over the {\tt sphere} model.


\begin{figure*}
\epsscale{1.0}
\plotone{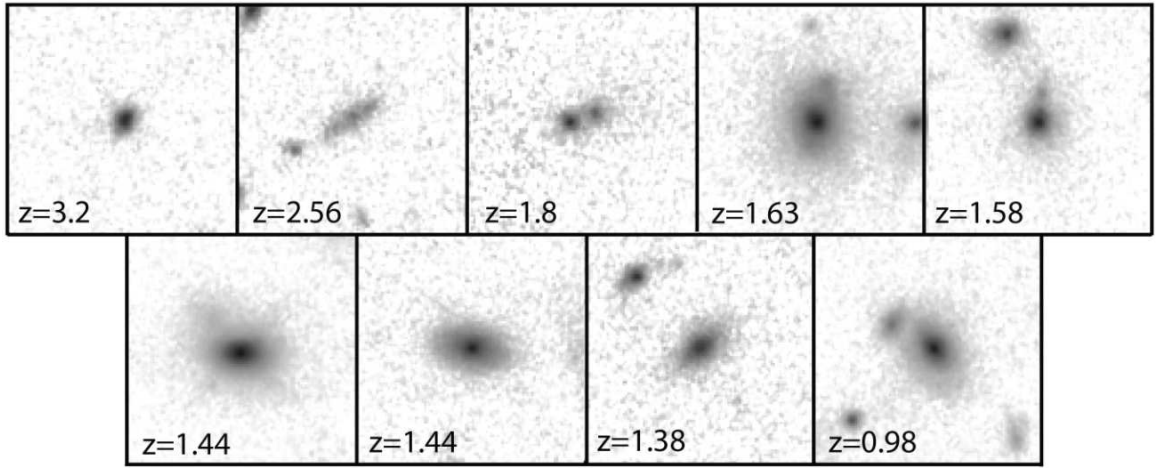}
\caption{\emph{Hubble}/WFC3 images taken in the F160W ($H$) band of the nine secure CTAGN candidates that fall within the CANDELS mosaic of the UDS field.  These sources have uncertainties on their $N_{\rm H}$ values such that $N_{\rm H}>10^{23.5}$ cm$^{-2}$ at the 90\% confidence level. A large fraction (six of nine) show a close companion or other morphological disturbance, in general agreement with the findings of Kocevski et al.~(2015).  \label{fig-ctagn-thumbs}}
\end{figure*}

For 13 sources that have more than 600 counts in their spectra, we allowed $\Gamma$ to be a free parameter. Furthermore, for 95 sources with less than 600 counts, where $\Gamma$ as a free parameter improves the fit ($\Delta C$-stat$\geq2.7$), we also allow $\Gamma$ to be a free parameter, unless $\Gamma\leq1.4$ in which case we keep it fixed at 1.7.

\begin{figure}
\epsscale{1.1}
\plotone{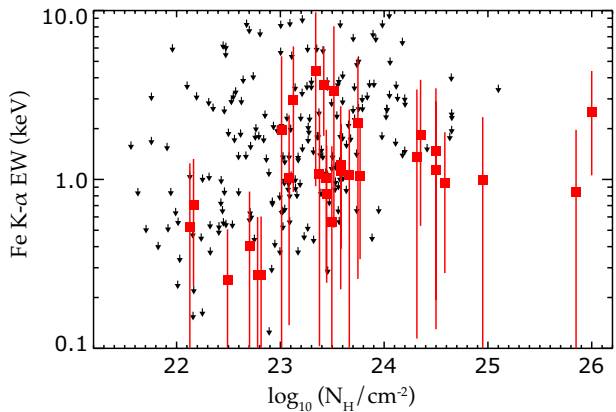}
\caption{Equivalent width of the Fe K-$\alpha$ line against $N_{\rm H}$. Red data points show sources where we could place a meaningful constraint on the line, whereas arrows show where only an upper limit could be determined. \label{fig_nh_ewfeka}}
\end{figure}

In Table 6 we provide the best-fitting spectral parameters, $N_{\rm H}$ and $\Gamma$, as well as their 90\% uncertainties, along with the observed 0.5$-$8 keV fluxes and absorption-corrected 2$-$10 keV rest-frame luminosities as determined from the spectral fit.  A more detailed description of each column in Table 6 is reported below.


\begin{enumerate}
\item Column 1: Source identification number
\item Column 2: Total (source + background) counts extracted
\item Column 3: Redshift
\item Column 4: Best-fit neutral hydrogen column density, $N_{\rm H}$, in units of log$_{10}(N_{\rm H}/{\rm cm}^{-2})$, with associated 90\% limits.  Errors of ``u'' and ``l'' denote that the uncertainty calculation hit the hard lower and upper limits of the model, which are $10^{20}$ cm$^{-2}$ and $10^{26}$ cm$^{-2}$, respectively. 
\item Column 5: Best-fit power-law index $\Gamma$ and associated 90\% limits (except where $\Gamma$ is fixed to 1.7; see text).  Errors of ``u'' and ``l'' denote that the uncertainty calculation hit the hard lower and upper limits of the model, which are 1.0 and 3.0, respectively.
\item Column 6: Observed $0.5-8$ keV fluxes in units of log$_{10}(F_{\rm X}/{\rm erg~cm}^{-2}~{\rm s}^{-1})$.
\item Column 7: Absorption-corrected, $2-10$ keV rest-frame luminosity in units of log$_{10}(L_{\rm X}/{\rm erg~s}^{-1})$.
\end{enumerate}


Figure \ref{fig-NHdist} presents histograms of the observed distributions of $N_{\rm H}$, flux, and luminosity that result from our spectral fitting.  Out of the 868 sources, we identify 51 Compton-thick AGN candidates, where $N_{\rm H}>10^{24}$ cm$^{-2}$.  Of these 51 sources, 29 have uncertainties on their $N_{\rm H}$ such that $N_{\rm H}>10^{23.5}$ cm$^{-2}$ at 90\% confidence.  Only seven sources have the statistics to constrain $N_{\rm H}>10^{24}$ cm$^{-2}$ at 90\% confidence.

Figure \ref{fig-xspec} shows the X-ray spectra of three of the Compton-thick AGN candidates detected in our sample.   All three sources exhibit strong Fe K$\alpha$ emission characteristic of a Compton-thick AGN.  A total of nine CTAGN candidates in our $>90$\% confidence subsample fall within the region of the UDS that has high-resolution near-infrared imaging from the CANDELS survey.  Thumbnail images of their host galaxies taken with \emph{Hubble}/WFC3 in the F160W filter ($H$ band) are shown in Figure \ref{fig-ctagn-thumbs}.  Interestingly, six of nine show a close companion or other morphological disturbance, in agreement with the findings of Kocevski et al.~(2015), who found an elevated merger fraction among the CTAGN relative to an unobscured control sample.  A more detailed analysis of the morphology of these host galaxies will be presented in a future paper.


We tested our results on the Compton-thick candidates by using the Murphy \& Yaqoob (2009) torus model ({\tt mytorus}) in place of the BN11 {\tt torus} model. The two models assume different geometries for the obscuring medium. We use {\tt mytorus} in coupled mode, where the parameters of the direct and scattered components are linked. We find 49 Compton-thick candidates with this model compared to 51 for the BN11 model, and in general very good agreement between the best-fit $N_{\rm H}$ values. Most of the discrepancies come from the model upper limit in {\tt mytorus}, which is $10^{25}$ cm$^{-2}$, rather than $10^{26}$ cm$^{-2}$ for {\tt torus}.


\begin{figure}
\epsscale{1.15}
\plotone{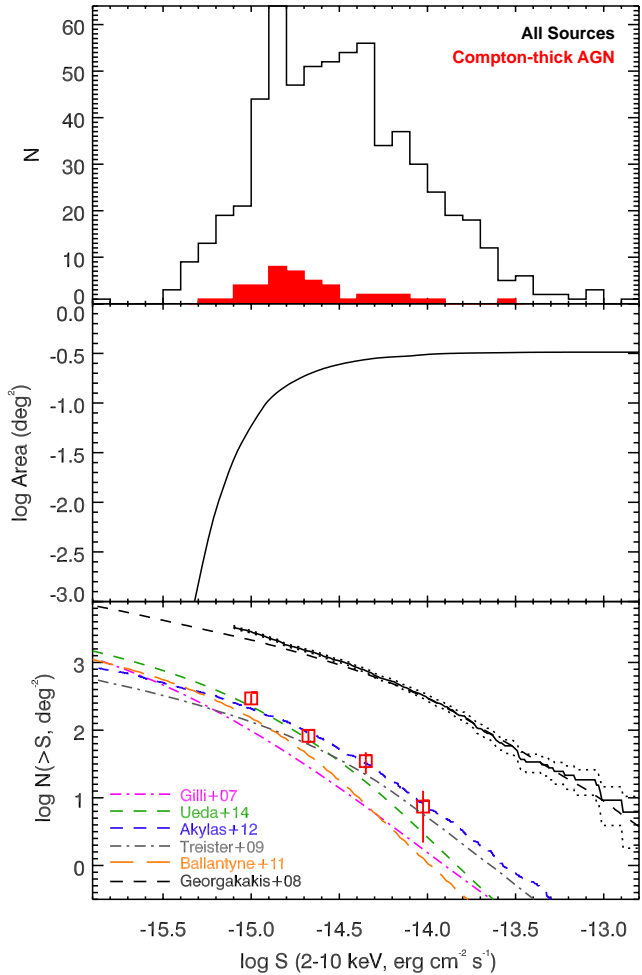}
\caption{(\emph{Top}) The distribution of 2-10 keV fluxes for sources in the X-UDS (black open histogram) and those identified as Compton-thick AGN (red solid histogram). (\emph{Middle}) Area of the X-UDS survey in the 2-10 keV band as a function of sensitivity. (\emph{Bottom}) Cumulative number counts, $N$, for all sources in the X-UDS as a function of sensitivity, $S$ (solid black line), with Poisson uncertainties (dotted black lines), compared to those presented in Georgakakis et al.~(2008, dashed black line). Number counts for Compton-thick AGN (red data points), also with Poisson uncertainties, are compared to predictions from several CXB synthesis models.}
\label{lognlogs}
\vspace*{0.1in}
\end{figure}

Finally, we investigated what constraints can be placed on the equivalent width (EW) of the Fe K-$\alpha$ line, which is strongly correlated with $N_{\rm H}$, where Compton-thick AGN are expected to have EW $>1$ keV.  Since the torus models include the line self-consistently, we replaced the torus model with {\tt pexrav} and {\tt plcabs} to model the continuum, plus a {\tt zgauss} component to model the line, and refit the spectra in order to calculate the EW.  We find, however, due to the low-count statistics of these sources, meaningful constraints on the Fe K-$\alpha$ line (i.e., normalization constrained to be $>0$) could only be placed on eight of our CT candidates. These all have EW $\sim$ keV as expected. Figure \ref{fig_nh_ewfeka} shows the measured EW values against the measured $N_{\rm H}$.  Arrows indicate where only an upper limit on the EW could be obtained.

Interestingly, we also identify high-EW sources among sources that do not otherwise show absorption in their X-ray spectra.  We find five such sources where we can constrain the EW to $> 0.3$ keV at 90\% confidence.  Monte Carlo simulations show that these EWs are challenging to produce for unobscured sight lines, even with Compton-thick material out of the line of sight (e.g., Brightman \& Nandra 2011).  To determine the significance of the Fe line feature in these sources, we ran spectral simulations using {\sc xspec}.  We simulated 1000 spectra of each source where the Fe line has been identified based on the best-fit unabsorbed power law, but without the Fe line. For each simulated spectrum, we determined the improvement in the fit statistic given the addition of a Gaussian line at the energy of redshifted Fe K$\alpha$.  We then count the number of simulated spectra where the improvement is as good as or better than the real spectrum.  We find that for four of the five sources identified above, only three to six out of 1000 simulated spectra satisfy this criterion. This represents a $\sim3\sigma$ detection of the line. These sources are xuds\_291, xuds\_524, xuds\_534 and xuds\_721.  The best-fit photon indices of these four sources are 1.51$^{+0.50}_{-0.47}$, 1.93$^{+0.96}_{-0.76}$, 1.93$^{+1.50}_{-1.12}$, 1.38$^{+1.34}_{-1.34}$, respectively.  These sources may also be CT candidates where both the reflected and transmitted components have been suppressed such that they are not evident in the {\it Chandra} spectra.  This might indicate extreme absorption above $10^{25}$ cm$^{-2}$.

\section{The Compton-thick fraction}

With our sample of 51 Compton-thick candidates (not including the high-EW sources), we proceed to calculate the intrinsic Compton-thick fraction. To do so, we follow the method of Brightman \& Ueda (2012), where the Compton-thick fraction for sources identified in the {\it Chandra} Deep Field South was presented. This requires accounting for several survey biases. These include the success rate of identifying CTAGN through spectral fitting and the contamination rate of sources identified as CT but whose true $N_{\rm H}$ is lower.  These are both ascertained through spectral simulations in {\sc xpec}.

For each source in the X-UDS catalog, with its count rate, background, and exposure time, we simulate 100 spectra based on the source's best-fit model.  We then refit the simulated spectra in the same way as above and determine either the percentage success rate for identifying it as Compton thick if the source was originally identified as CT, or the contamination rate if the original source was not identified as Compton thick. The typical success rate is 60\%, whereas the contamination rate is low, around 1\%. When calculating the Compton-thick fraction, we normalize by the success rate and subtract the expected number of contaminating sources.

We then calculate the inferred number density for each source based on its count rate and the survey area curve from Figure 5. This is important since the CTAGN have lower fluxes than their unobscured counterparts, and at lower fluxes the survey area is smaller, biasing their number counts downward. We show this in Figure \ref{lognlogs} which shows a histogram of 2-10 keV observed fluxes for all the X-UDS sources as well as the CTAGN we have identified. Also shown is the area curve for sources selected in the 2-10 keV band.

In Figure \ref{lognlogs} we show the cumulative number counts as a function of sensitivity where the uncertainty plotted is the Poisson uncertainty. For all sources, we compare the number counts to those presented in Georgakakis et al.~(2008).  This shows broad agreement over a wide range of sensitivities, although at sensitivities of $\sim10^{-15}$ ergs cm$^{-2}$ s$^{-1}$ some disparity can be seen, likely due to the small and declining area of the X-UDS survey at these fluxes and Eddington bias. Therefore, for the rest of our analysis we only consider sources with fluxes greater than $10^{-15}$ ergs cm$^{-2}$ s$^{-1}$.

Also shown in this Figure \ref{lognlogs} are the cumulative number counts for the CTAGN with their Poisson uncertainties, accounting for the success rate of CT identification and contamination described above.  We compare our results on the CTAGN number counts to several predictions of CXB synthesis models, including Gilli et al.~(2007), Akylas et al.~(2012, assuming a 40\% CT fraction), Ueda et al.~(2014) and the model from Ballantyne et al.~(2011), adapted for Harrison et al.~(2016).  Our data prefer models with higher CT number counts from Akylas et al.~(2012) and Ueda et al.~(2014).

\begin{figure}
\epsscale{1.15}
\plotone{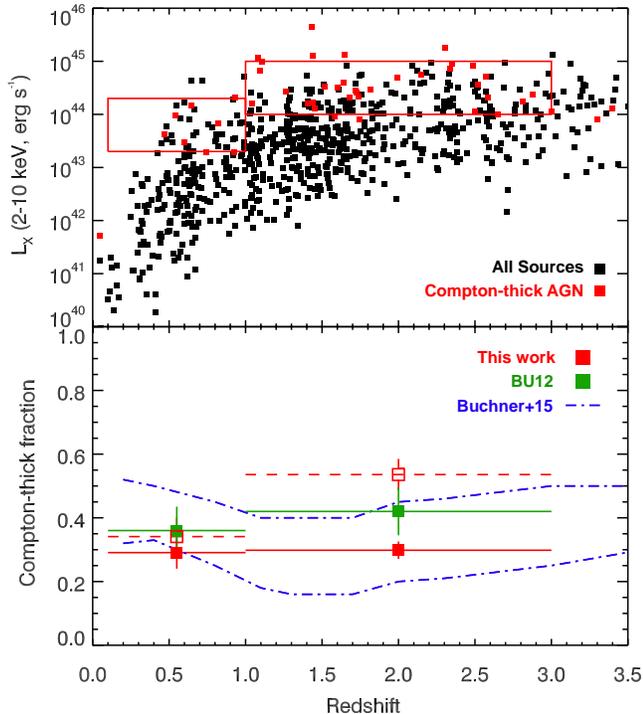}
\caption{(\emph{Top}) The distribution of intrinsic (unabsorbed) 2-10 keV luminosities and redshifts for sources in the X-UDS (black squares) and those identified as Compton-thick AGN (red squares). Red boxes show the redshift and luminosity ranges where we evaluate the Compton-thick fraction. (\emph{Bottom}) The derived Compton-thick fraction as a function of redshift (red solid squares) and the fraction normalized to a luminosity of $10^{43.5}$ erg s$^{-1}$ (red open squares, assuming a luminosity dependence of the fraction). The uncertainties on the Compton-thick fraction have been propagated from the Poisson errors on $N(N_{\rm H}>10^{24}$ cm$^{-2})$ and $N(N_{\rm H}<10^{24}$ cm$^{-2})$. We compare our results to previous work from Brightman \& Ueda (2012) and Buchner et al.~(2015, plotted as 90\% confidence interval), both evaluated at $L_{\rm X}=10^{43.5}$ erg s$^{-1}$.}
\vspace*{0.1in}
\label{ctfrac}
\end{figure}

Finally, we calculate the intrinsic CT fraction. Since our sources cover a wide range of redshifts, we calculate the Compton-thick fraction as a function of epoch. Figure \ref{ctfrac} shows how the X-UDS sources are distributed in redshift and 2-10 keV luminosity and also shows the CTAGN we have identified. Although the observed fluxes of these sources are typically low, their intrinsic luminosities, determined though spectral fitting, are relatively high compared to the other sources in the survey because they have been corrected upward by several factors. 

In order to calculate the CT fraction in an unbiased way, we have to define regions where the survey is complete to all sources given a redshift bin.  We show these as red boxes in Figure \ref{ctfrac}.  While these are arbitrary, we do not find that our results depend strongly on the exact value of the bin limits.  We calculate the CT fraction for each redshift bin from the number densities, corrected for the biases as mentioned above.  Uncertainties on the CT fraction are calculated by propagating the Poisson errors on the number of AGN with $N_{\rm H}$ above and below $10^{24}$ cm$^{-2}$.  Our derived CT fractions are shown in the bottom panel of Figure \ref{ctfrac}, where we also compare to previous results from Brightman \& Ueda (2012) and Buchner et al. (2015).  Since the Compton-thick fraction may be luminosity dependent, we normalize the Compton-thick fraction to a common luminosity of $10^{43.5}$ erg s$^{-1}$ as described in Brightman \& Ueda (2012).  

In our lowest redshift bin ($z=0.1-1$), our data indicate a Compton-thick fraction of $30\%-35\%$ which is consistent with the latest estimate of the CT fraction in the local universe from Ricci et al.~(2015), and with the results from Buchner et al.~(2015) for a large range of redshifts.  We find a similar fraction at higher redshifts ($z=1-3$), although this is for higher luminosities.  This would support a Compton-thick fraction that does not vary strongly with luminosity or redshift, as suggested by Buchner et al.~(2015).  However, assuming a luminosity dependence of the CT fraction, which is well known for less obscured sources, our results would be consistent with a rise in the CT fraction with increasing redshift.


\section{Summary and Future Work}

We have introduced the X-UDS survey, the first deep \emph{Chandra} observations of the Subaru-XMM Deep/UKIDSS UDS field.  As an X-ray Visionary Project, X-UDS is designed to facilitate a diverse set of science goals, ranging from uncovering the nature of the first accreting SMBHs at $z>6$ to studying the prevalence and demographics of heavily obscured, Compton-thick AGN out to $z\sim2$.  The survey covers a total area of 0.33 deg$^{2}$ and the ACIS-I observations have been tiled to provide a nominal depth of $\sim600$ ksec in the central region of the field that has been imaged by the CANDELS survey and $\sim200$ ksec in the remainder of the field.

We have presented a catalog of 868 band-merged point sources detected with a false-positive Poisson probability of $<1\times10^{-4}$.  Sensitivity maps and sensitivity curves were made following the method of Georgakakis et al.~(2008), which accounts for the observational biases that affect the probability of detecting a source at a given flux.  We estimate our flux limits to be $4.4\times10^{-16}$ (0.5-10 keV), $1.4\times10^{-16}$ (0.5-2 keV), $6.5\times10^{-16}$ (2-10 keV), and $9.2\times10^{-16}$ (5-10 keV) erg cm$^{-2}$ s$^{-1}$ in our four analysis bands.  Based on single-band detections in the $4-7$ keV energy range, we have determined that 36 sources may be spurious over our four detection bands, placing our spurious detection fraction at 4.1\%.

In addition, we have carried out a spectral analysis on a subsample of 457 sources detected with greater than 20 counts in the full band following the methodology of Brightman et al.~(2014).  We provide best-fitting spectral parameters $N_{\rm H}$ and $\Gamma$, as well as the absorption-corrected 2$-$10 keV rest-frame luminosities for each source.  We present a sample of 51 Compton-thick AGN candidates that have nuclear obscuration values of $N_{\rm H}>10^{24}$ cm$^{-2}$ as determined by our spectral fits.

Based on our sample of heavily obscured AGN, we estimate the intrinsic Compton-thick fraction to be 30-35\% at both low ($z=0.1-1$) and high ($z=1-3$) redshifts.  This fraction does not vary strongly with luminosity or redshift, in agreement with the findings of Buchner et al.~(2015).  However, if we assume a luminosity dependence of the Compton-thick fraction and normalize our results to a common luminosity, our results do support a rise in the Compton-thick fraction with increasing redshift.  Finally, the hosts of nine of our Compton-thick AGN show a high fraction of close companions or morphological disturbances, in agreement with the results of Kocevski et al.~(2015).


We anticipate several future papers on a wide range of science topics given the wealth of multiwavelength ancillary data in the UDS field.  This includes work on optical counterpart matching, an analysis of diffuse detections in the field, and a more thorough examination of AGN host morphologies as a function of their nuclear obscuration.  However, the power of the X-UDS dataset will be fully realized when used in conjunction with other legacy surveys.  X-UDS effectively completes \emph{Chandra's} observations of the five premier extragalactic survey fields.  Together, this combined dataset offers the most comprehensive census of AGN ever compiled at moderate to high redshifts and will facilitate the study of thousands of AGN and their host galaxies over the redshift, luminosity, and column-density range responsible for the bulk of SMBH growth in the Universe.


All of the data products described in this paper, including event files, exposure maps ,and catalogs, are available via the public website http://www.mpe.mpg.de/XraySurveys/.  The X-UDS dataset has also been implemented into the web-based stacking analysis tool CSTACK\footnote{http://lambic.astrosen.unam.mx/cstack}. The X-UDS implementation of CSTACK will be publicly available one year after the publication of this paper.

\vspace{0.25in}  
We acknowledge the financial support from NASA \emph{Chandra} grant GO5-16150B (D.D.K, G.H.).  TM is supported by UNAM-DGAPA PAPIIT IN104216 and CONACyT 252531.  MB acknowledges support from the FP7 Career Integration Grant ``eEASy’' (CIG 321913). 

\bibliography{}

\end{document}